\documentstyle[preprint,aps]{revtex}

\begin{document}

\tighten
\draft
\preprint{
\vbox{
\hbox{November 1994}
\hbox{TPR--94--34}
\hbox{ADP--94--22/T162}
}}

\title{Spin-dependent nuclear structure functions:
       general approach with application to the Deuteron
       \footnote{Work supported in part by BMFT grant 06 OR 744(1)}}
\author{S.A.Kulagin$^a$
        \footnote{On leave from the Institute for Nuclear Research
                  of the Russian Academy of Sciences,
                  60th October Anniversary Pr. 7a,
                  117312 Moscow, Russia},
        W.Melnitchouk$^{a,b}$, G.Piller$^c$ and W.Weise$^{a,b}$}
\address{$^a$ Institut f\"{u}r Theoretische Physik,
         Universit\"{a}t Regensburg,
         D-93040 Regensburg, Germany.}
\address{$^b$ Physik Department,
         Technische Universit\"{a}t M\"unchen,
         D-85747 Garching, Germany.}
\address{$^c$ Department of Physics and Mathematical Physics,
         University of Adelaide,
         S.A. 5005, Australia.}

\maketitle

\begin{abstract}
We study deep-inelastic scattering from polarized
nuclei within a covariant framework.
A clear connection is established between
relativistic and non-relativistic limits,
which enables a rigorous derivation of convolution
formulae for the spin-dependent nuclear structure
functions $g_1^A$ and $g_2^A$
in terms of off-mass-shell extrapolations
of polarized nucleon structure functions,
$g_1^N$ and $g_2^N$.
Approximate expressions for $g_{1,2}^A$ are obtained
by expanding the off-shell $g_{1,2}^N$ about their
on-shell limits.
As an application of the formalism we consider
nuclear effects in the deuteron, knowledge
of which is necessary to obtain accurate information
on the spin-dependent structure functions of the neutron.
\end{abstract}
\pacs{PACS numbers: 13.60.Hb, 13.88.+e, 24.70.+s}

\section{Introduction}
\label{introdu}

Polarized deep-inelastic scattering (DIS) experiments have
in recent years yielded a number of important and sometimes
unexpected results.
The measurements by the European Muon Collaboration (EMC)
of the $g_1^p$ structure function of the proton \cite{EMC}
over a large range of values of the Bjorken scaling variable $x$,
when combined with flavor non-singlet matrix elements from weak
decays, provided information on the singlet axial charge
of the proton.
The small size of this resulted in the so-called proton
``spin crisis'', which prompted a serious reanalysis of the very
ideas behind the quark model and the simple parton picture of DIS.
More recent experiments on the proton by the
Spin Muon Collaboration
(SMC) \cite{SMCP} and the SLAC E143 Collaboration \cite{E143P}
have allowed more refined analyses of the $x$ and $Q^2$ dependence
of the $g_1^p$ structure function \cite{EK,CR,ANR}.
(For a recent review see Ref.\cite{BLOO}.)

As an extra source of information, it is important also
to measure the neutron polarized structure function,
$g_1^n$.
Besides revealing the spin structure of the neutron itself,
measurement of $g_1^n$ is essential for testing the fundamental
Bjorken sum rule.
The absence of free neutron targets means, however, that
light nuclei have to be used instead for this purpose.
The SLAC E142 Collaboration \cite{E142} has in fact measured
the structure function of ${}^3$He, which, because of the
preferential antiparallel polarization of protons in
the ${}^3$He nucleus, is believed to be approximately equal
to the polarized structure function of the neutron.
In addition, the SMC \cite{SMCD} has recently measured the
$g_1$ structure function of the deuteron --- combined with
either the proton or neutron (${}^3$He) data, this can
be used as a valuable consistency check on the other
measurements.

To obtain accurate information on nucleon structure functions
from nuclear DIS data, it is of course essential to reliably
subtract any nuclear effects in the extraction procedure.
Away from the shadowing region at small Bjorken $x$ ($x \alt 0.1$),
the standard method for investigating nuclear effects is the
so-called convolution model, in which the nuclear structure
function is expressed as a (one-dimensional) convolution of
the spin-dependent nucleon structure function and the nucleon
momentum distribution in the nucleus.
The convolution model follows from the impulse approximation,
Fig.\ref{F_DIS}, if one assumes that factorization between
photon--nucleon and nucleon--nucleus scattering amplitudes
translates into factorization between structure functions,
although in a covariant framework this assumption is
generally not justified.

For ${}^3$He targets, nuclear effects have been investigated
in Refs.\cite{KAP1,CIOF}, and for the deuteron in
Refs.\cite{KAP1,FS,WOL,KAP2,TOK,MPT,UMN}.
In Refs.\cite{TOK,UMN} relativistic effects in the deuteron
were also included, although still within the confines
of the convolution model.
It was shown in Ref.\cite{MPT}, however, that relativistic
corrections necessarily lead to a breakdown of convolution
for $g_1^A$ when the full off-mass-shell structure of bound
nucleons is incorporated.
Although the convolution-breaking effects are not large
(typically $\sim 0.5\%$ in the deuteron \cite{MPT}),
in any self-consistent calculation they must be included.

Aside from the relativistic complications,
the situation is not completely clear
even in the non-relativistic approaches.
There exists in the literature \cite{KAP1,CIOF,FS,WOL,KAP2}
a variety of results for convolution formulae for $g_1^A$,
the derivation of which is often based on early convolution
models for unpolarized scattering \cite{CONV}, in which the
issue of off-shell effects was not seriously addressed.
The need exists, therefore, to derive convolution formulae
for spin-dependent structure functions systematically in
the non-relativistic limit.

In this paper we present an analysis of the polarized
structure functions of nuclei, starting from a covariant
framework, and working consistently to order ${\bf p}^2/M^2$
in the bound nucleon momentum.
We demonstrate that in this limit one does indeed recover
(two-dimensional) convolution formulae, although with
different ``flux factors''
(polarized nucleon momentum distributions)
compared to those found in the literature.
Our formalism enables us to consider both the $g_1$ and $g_2$
structure functions (or equivalently the transverse structure
function $g_T \equiv g_1 + g_2$) on a similar footing.
For the latter, we find that $g_2^A$ receives contributions
from the $g_1^N$ as well as from the $g_2^N$ structure functions
of the nucleon.
All of the formal results are valid in the Bjorken limit
for spin 1/2 and spin 1 nuclei.

Taking advantage of the weak binding of nucleons in nuclei,
we also derive expansion formulae for $g_{1,T}^A$ in terms
of derivatives of $g_{1,T}^N$.
We concentrate on the specific case of the deuteron,
where we present a detailed comparison of the expansion
formula results with those of the full non-relativistic
convolution, as well as with previous relativistic
calculations.
In addition, we estimate for the first time the
nuclear effects that one needs to account for when
separating the nucleon $g_T$ (or $g_2$) structure
function from deuteron data.
This will be important in view of upcoming experiments
which will for the first time measure $g_2$
for deuterium targets.

The contents of this paper are laid out as follows:
in Section~\ref{framework} we present the general covariant framework
in which $g_{1,T}^A$ of a nucleus $A$ are expressed
in terms of the off-shell nucleon propagator and the
virtual nucleon hadronic tensor;
in Section~\ref{nr-limit} we describe how the relativistic expressions
can be reduced by making a non-relativistic expansion of the
nucleon propagator in medium;
Section~\ref{expansion} deals with the details of approximate expansion
formulae which are obtained by Taylor-expanding the off-shell
nucleon structure function about its on-shell limit;
application of the formalism to the case of deuterium
is presented in Section~\ref{deuteron};
finally concluding remarks are made in Section~\ref{finita}.

\section{Relativistic Framework}
\label{framework}

\subsection{Definitions}

Inclusive deep-inelastic scattering of leptons from hadrons
is described by the hadronic tensor
\begin{eqnarray}
W_{\mu \nu}(P,q,S)
&=& \frac{1}{2\pi} \int d^4\xi\, e^{iq\cdot \xi}
\left\langle P,S \left| \left[ J_{\mu}(\xi), J_{\nu}(0)
                        \right]
                 \right|P,S
\right\rangle,
\end{eqnarray}
where $P$ and $q$ are the four-momenta of the target and photon,
respectively, and the vector $S$ is orthogonal to the target
momentum,
$P\!\cdot\!S=0$, and normalized such that $S^2=-1$.
For spin-1/2 targets, $S$ is simply the target polarization vector,
while for the spin-1 case $S$ is defined in terms of
polarization vectors $\varepsilon_{\alpha}^m$ such that
$S^{\alpha}(m)
= -i \epsilon^{\alpha}
     \left( \varepsilon^{m *}, \varepsilon^{m}, P \right) / M_T$,
where $m=0,\pm1$ is the spin projection along
the axis of quantization,
$M_T$ denotes the target mass,
and we define $\epsilon^{\alpha}(a,b,c)\equiv
\epsilon^{\alpha\beta\mu\nu} a_{\beta} b_{\mu} c_{\nu}$.
The hadronic tensor can be decomposed into symmetric ($s$)
and antisymmetric ($a$) parts,
\begin{eqnarray}
W_{\mu\nu}(P,q,S)
&=& W_{\mu\nu}^{(s)}(P,q,S)\ +\ i\ W_{\mu\nu}^{(a)}(P,q,S).
\end{eqnarray}
With unpolarized (charged) lepton beams one is sensitive only
to the symmetric part, which, at leading twist, depends on
the spin-independent $F_1$ and $F_2$ structure functions,
and, for spin-1 targets, also on the structure functions
$b_{1,2}$ when the target alone is polarized \cite{HJM}.
If the lepton and hadron are both polarized,
only the antisymmetric component of $W_{\mu\nu}$
is relevant.
For either spin-1/2 or spin-1 targets this is expressed
in terms of the two independent, dimensionless structure
functions $g_{1,2}$:
\begin{eqnarray} \label{Wa}
W_{\mu\nu}^{(a)}(P,q,S)
&=& \frac{1}{2P\!\cdot\!q}
    \epsilon_{\mu\nu\alpha\beta}\,q^{\alpha}\, G^{\beta}(P,q,S), \\
G^{\beta}(P,q,S)
&=& 2M_T
    \left[S^{\beta}\left(g_1 + g_2 \right) -
    P^{\beta}\frac{S\!\cdot\!q}{P\!\cdot\!q} g_2 \right].
\end{eqnarray}
In the Bjorken limit ($Q^2, P\cdot q \rightarrow \infty$),
in which we work throughout, both structure functions
$g_1$ and $g_2$ exhibit scaling,
i.e. up to logarithmic QCD corrections they depend only
on the ratio $Q^2/2P\cdot q$.

\subsection{Nucleon Tensor and Structure Functions}

In our covariant analysis it will be useful to work with the
off-shell nucleon tensor $\widehat {\cal W}_{\mu\nu}$,
which is defined through the imaginary part of the forward photon
scattering amplitude from an
off-shell nucleon.
In terms of $\widehat {\cal W}_{\mu\nu}$, the hadronic tensor of an
on-shell nucleon ($p^2 = M^2$ where $p$ is the nucleon momentum) is:
\begin{eqnarray}\label{What}
W_{\mu\nu}^N(p,q,s)
&=& \text{Tr} \left[
(\overlay{\slash}p+M)
        \frac{(1+\gamma_5\overlay{\slash}s)}{2}\
                        \widehat {\cal W}_{\mu\nu}(p,q) \right],
\end{eqnarray}
where $s$ is the nucleon spin vector, and $M$ is the nucleon mass.
The antisymmetric part of $\widehat {\cal W}_{\mu\nu}$
is given by an expression similar to that in Eq.(\ref{Wa}),
namely:
\begin{eqnarray}
\widehat {\cal W}_{\mu\nu}(p,q)
&=& \frac{1}{2p\cdot q}
\epsilon_{\mu\nu\alpha\beta}\,q^{\alpha}\,
\widehat{\cal G}^{\beta}(p,q).
\label{Whatdef}
\end{eqnarray}

In analyzing the tensor structure of $\widehat{{\cal W}}_{\mu\nu}$
it is convenient to expand $\widehat{\cal G}^{\beta}$
in terms of a complete set of Dirac matrices,
$ \left\{ I, \gamma^\alpha, \sigma^{\alpha\beta},
          \gamma^\alpha\gamma_5, \gamma_5
\right\}$.
The various coefficients in this expansion
must be constructed from the vectors $p$ and $q$,
and from the tensors $g_{\alpha\beta}$ and
$\epsilon_{\mu\nu\alpha\beta}$.
Terms proportional to $q^\beta$ do not contribute to
$\widehat{{\cal W}}_{\mu\nu}$, and are therefore not considered.
The requirements of parity, time-reversal invariance and
hermiticity restrict further the number of possible terms.
In particular, it follows from parity invariance that
$\widehat{{\cal G}}^\beta$ transforms as an axial vector
and the coefficient of the scalar ($I$) term must be zero.
The requirements of time reversal invariance and hermiticity
rule out the vector ($\gamma^\alpha$) and the pseudoscalar
($\gamma_5$) terms as well.
Finally we find that $\widehat{{\cal G}}^\beta$
can be written in terms of six independent structures:
\begin{eqnarray}
\widehat{\cal G}^{\beta}(p,q)
&=& \left(
          \frac{G^{(p)}}{M^2}\overlay{\slash}p
         + \frac{G^{(q)}}{p\cdot q}\overlay{\slash}q
    \right)\, p^{\beta}\gamma_5
 + G^{(\gamma)} \gamma^{\beta}\gamma_5 \nonumber\\
&+& \left( { G^{(\sigma p)} \over M }\, p_{\alpha}
         + { G^{(\sigma q)} \over p\cdot q }\, q_{\alpha}
    \right) i\gamma_5 \sigma^{\beta\alpha}
 + { G^{(\sigma p q)} \over M\ p\cdot q }\,
    p^{\beta} i \gamma_5 \sigma^{\alpha\lambda}\,
    p_{\alpha} q_{\lambda},
\label{DirSt}
\end{eqnarray}
where the coefficient functions $G^{(i)}$ are constructed
to be scalar, dimensionless and real functions of
$q^2$, $p\cdot q$ and $p^2$.

Substituting $\widehat{\cal G}^{\beta}$ into Eq.(\ref{What})
we can express the on-shell nucleon
\footnote{Throughout we define $g_{1,2}^N$ to be
the average of the proton and neutron structure functions:
$g_{1,2}^N = \left. \left( g_{1,2}^p\ +\ g_{1,2}^n \right)
             \right/2$.}
structure functions $g^N_{1,2}$ in terms of the
coefficients $G^{(i)}$ (see also Ref.\cite{MPT}):
\begin{mathletters}
\label{g12on}
\begin{eqnarray}
g_1^N
&=&  G^{(q)} + G^{(\gamma)} + G^{(\sigma p)} - G^{(\sigma pq)},  \\
g_2^N
&=& -G^{(q)} + G^{(\sigma q)} + G^{(\sigma pq)},
\end{eqnarray}
\end{mathletters}%
with the functions $G^{(i)}$ evaluated here at
their on-shell values.
Note that the term proportional to $\overlay{\slash}p\gamma_5$
vanishes when tracing $\widehat{\cal G}^{\beta}$
with the projection operator in Eq.(\ref{What}),
so that only five terms out of a possible six in Eq.(\ref{DirSt})
contribute to the physical nucleon structure functions.
The structure $\overlay{\slash}p\gamma_5$ could,
in general, give non-vanishing contribution to structure functions
of nuclei.
However, as we shall see in Section~\ref{nr-limit},
only the above five
functions will be relevant in the non-relativistic limit.

\subsection{Nuclear Structure Functions}

Discussions of nuclear effects in deep-inelastic scattering
are usually framed within the context of the impulse approximation
for the nucleons, see Fig.\ref{F_DIS}.
Other possible nuclear effects which go beyond the impulse
approximation are final state interactions between the
recoiling nucleus and the debris of the struck nucleon
\cite{SAIT}, corrections due to mesonic exchange currents
\cite{KUL,KAP3,MT}, and nuclear shadowing
\footnote{Some potential problems associated with the use
of the impulse approximation for the $g_2$ structure function
have also been discussed in Ref.\cite{MS} in the context of
relativistic light-front dynamics.}.
One may argue that complications due to meson exchange
currents are less important here than in unpolarized scattering
since their main contribution comes from pion exchange.
Because it has spin zero, direct scattering from a pion constituent
of a nucleus gives no contribution to spin-dependent structure
functions.
Also coherent multiple scattering effects, which are known to lead
to nuclear shadowing, should not be important for large values
of the nucleon Bjorken scaling variable $x = Q^2/2Mq_0$.
This is evident if one recognizes that the characteristic time
scale $1/Mx$ of the DIS process is smaller than the typical average
distance between bound nucleons in the nucleus for $x > 0.1$.
Based on these observations we consider the diagram
in Fig.\ref{F_DIS} as a basic approximation.
In this case the nuclear hadronic tensor
can be written:
\begin{eqnarray}
\label{IA}
W_{\mu\nu}^A(P,q,S)
= \int [dp]\ \text{Tr} \left[
  {\cal A}(p;P,S)\ \widehat{\cal W}_{\mu\nu}(p,q)
  \right],
\end{eqnarray}
with $\left[dp\right] \equiv d^4p / (2\pi)^4i$.
The function ${\cal A}(p;P,S)$ is the nucleon propagator inside
the nucleus with momentum $P$ and polarization $S$,
\begin{eqnarray}
\label{A}
{\cal A}(p;P,S)
&=& -i\int d^4\xi\,e^{ip\cdot \xi}
\left\langle
P,S\left|T \left( N(\xi)\overline{N}(0) \right) \right|P,S
\right\rangle.
\end{eqnarray}
Here $N(\xi)$ is the nucleon field operator,
and $\widehat{\cal W}_{\mu\nu}(p,q)$ is the hadronic tensor
of the off-mass-shell nucleon, given by Eqs.(\ref{Whatdef})
and (\ref{DirSt}).

The expression for the nuclear tensor in Eq.(\ref{IA})
is covariant and can be evaluated in any frame.
It will be convenient, however, to work in the target rest frame,
in which the target momentum is $P=(M_A;{\bf 0})$,
and the momentum transfer to the nucleus,
$q=(q_0;{\bf 0}_\perp,-|{\bf q}|)$, defines the $z$-axis.
For the $g_1$ structure function it is natural to choose
the spin quantization axis such that the target is polarized
in a direction longitudinal to the momentum transfer,
$S=S^{\|}=(0;{\bf 0}_\perp,1)$.
Taking the $W^A_{12}$ component in Eq.(\ref{IA}), which is
proportional to $g_1^A$, we find:
\begin{mathletters}
\label{g1TA}
\begin{eqnarray}
x_A g_1^A(x_A)
&=& \frac1{2M_A}\int [dp]\ x'\ \text{Tr}
    \left[
    {\cal A}^{\|}(p)\,
    \left(\widehat{\cal G}_0 + \widehat{\cal G}_z \right)
    \right],
\label{g1A}
\end{eqnarray}
where $x_A=Q^2/2P \cdot q$ and $x'=Q^2/2p \cdot q$ are the
Bjorken variables for the nucleus and bound (off-mass-shell)
nucleon, respectively.
The nucleon propagator in the nucleus at rest with polarization
$S^{\|}$ is denoted by ${\cal A}^{\|}(p)={\cal A}(p;S^{\|})$.

The transverse spin-dependent structure function of the nucleus,
$g_T^A \equiv g_1^A + g_2^A$, is obtained by choosing the target
polarization in a direction perpendicular to the momentum transfer,
$S=S^{\perp}=(0;{\bf S}^\perp,0)$,
and taking the $W^A_{13}$ and $W^A_{23}$ components
in Eq.(\ref{IA}):
\begin{eqnarray}
x_A g_T^A(x_A)
&=& \frac1{2M_A}\int [dp]\ x'\ \text{Tr}
     \left[
     {\cal A}^{\perp}(p)\, \widehat{\cal G}_\perp
     \right],
\label{gTA}
\end{eqnarray}
\end{mathletters}%
where ${\cal A}^{\perp}(p) = {\cal A}(p;S^{\perp})$,
and $\widehat{\cal G}_\perp$ represents the transverse
spatial components of $\widehat{\cal G}^{\beta}$,
corresponding to the transverse quantization axis.
{}From Eqs.(\ref{g1A}) and (\ref{gTA}) one can
reconstruct the $g_2^A$ structure function by taking
the difference between $g_T^A$ and $g_1^A$.

We should stress that the treatment culminating in the
results of Eqs.(\ref{g1TA}) has been fully relativistic,
and exact within the impulse approximation.
In the literature one usually encounters formulations
in terms of simple convolution formulae
\cite{KAP1,CIOF,FS,WOL,KAP2,TOK,UMN},
in which the nuclear structure functions are expressed
as one-dimensional convolutions of the nucleon momentum
distribution in the nucleus, and the (on-shell)
structure functions of the nucleon.
To obtain the simple convolution result one requires
two conditions to be satisfied: firstly, that the traces
${\rm Tr}\, [ {\cal A}\,\widehat{{\cal G}} ]$
in Eqs.(\ref{g1TA}) factorize into completely separate
nuclear and nucleon parts,
and secondly that the nucleon component be
independent of $p^2$.
For the twist-2 structure function $g_1^A$,
it was shown in Ref.\cite{MPT} that
in a relativistic treatment the off-shell
degrees of freedom associated with bound nucleons
in fact violate both conditions, leading to
a breakdown of the simple convolution picture.
A similar breakdown must also occur for the $g_2^A$ structure
function, since its twist-2 contribution contains a component
proportional to $g_1^N$ (see Section~\ref{sf-input} below).

Clearly, in a relativistic theory one needs to go beyond
the simple convolution formulation.
On the other hand, the advantage of the convolution model
is its ease of application.
In the study of DIS from nuclei, especially light nuclei
where typical binding energies are small,
it may in fact be quite sufficient to treat
the nucleus as a non-relativistic system.
It was shown in Ref.\cite{KPW} that
for unpolarized nuclear structure functions, in which both
of the above criteria are also violated relativistically
\cite{MST,MSTD}, one can in fact recover factorized
expressions in the non-relativistic limit.
It is possible to define ``off-shell nucleon structure
functions'', with the correct on-shell limits, which
lead to (two-dimensional) convolution formulae.
In the next section we perform a similar non-relativistic
reduction of the relativistic expressions in
Eqs.(\ref{g1TA}) to establish whether a similar factorization
is attainable in spin-dependent processes.

\section{Nuclear Structure Functions in the Non-Relativistic Limit}
\label{nr-limit}

Our basic assumption in the remainder of the paper is that the
nucleus is a non-relativistic system, made up of weakly bound
nucleons interacting via the exchange of mesonic fields.
This necessarily involves neglecting antinucleon degrees of freedom,
and corresponds to bound nucleons in the nucleus being slow,
$|{\bf p}|\ll M$.
With these assumptions we will derive from Eqs.(\ref{g1TA})
simplified (convolution) expressions for the spin-dependent
$g_{1,T}^A$ structure functions of weakly bound nuclei.

\subsection{Non-Relativistic Reduction}

Following the procedure outlined for example in Ref.\cite{LL},
we can derive the relation between the relativistic nucleon field
operator $N$ and the non-relativistic operator $\psi$.
A detailed discussion of the non-relativistic reduction of $N$
is given in Appendix~\ref{nr-reduction}.
The essential result is that, up to order
${\bf p}^3/M^3$ corrections,
the operators $N$ and $\psi$ are connected via:
\begin{equation}
\label{N-psi}
N({\bf r},t) = e^{-iMt}
\left(
        \begin{array}{r}
                Z\,\psi({\bf r},t) \\
\frac{
{\mbox{\boldmath$\sigma\cdot$}\displaystyle\bf p}
  }{\displaystyle 2M}\,\psi({\bf r},t)
        \end{array}
\right),
\end{equation}
where ${\bf p} \equiv -i${\boldmath$\nabla$},
and the renormalization operator $Z=1-{\bf p}^2/8M^2$
guarantees baryon number conservation.
This result is valid for a wide range of meson--nucleon
interactions, in particular for interactions with scalar
and vector mesons as well as for pseudovector coupling
to pions.
Furthermore, it is explicitly interaction--independent.
For the pseudoscalar $\pi N$ couplings discussed in
Refs.\cite{KAP3,RCT}, however, one finds explicit
interaction dependence in Eq.(\ref{N-psi}), although
the pseudoscalar interaction is generally considered
less reliable than the pseudovector model, which we
restrict ourselves to in this paper
(see Appendix~\ref{nr-reduction}).

Consider now the consequences of applying Eq.(\ref{N-psi})
to the traces in Eqs.(\ref{g1TA}).
We start by writing the nucleon propagator (\ref{A}) as:
\begin{eqnarray} \label{A2}
\frac1{2M_A}{\cal A}_{\alpha\beta}(p)
&=& -i\int\! dt\,e^{ip_0t}
\left\langle
T \left( N_{\alpha}({\bf p},t)\overline{N}_{\beta}({\bf p},0)
  \right)
\right\rangle,
\end{eqnarray}
where
$N_\alpha({\bf p},t)
= \int\! d^3{\bf r}\,e^{-i{\bf p\cdot r}}N_\alpha({\bf r},t)$
is the nucleon operator in a mixed $({\bf p},t)$-representation,
and the brackets denote an average over the nuclear state,
$\left\langle \cdots \right\rangle
\equiv \left\langle A|\cdots|A \right\rangle
   /   \left\langle A|A \right\rangle$.
Writing the four-momentum of the bound nucleon as
$p=(M+\varepsilon; {\bf p})$,
we can introduce the non-relativistic nucleon propagator
${\cal A}^{\rm NR}$, which is usually defined as:
\begin{equation}
\label{ANR}
{\cal A}^{\rm NR}_{\sigma\sigma'}(\varepsilon,{\bf p})
= -i\int dt\,
    e^{i\varepsilon t}
    \left\langle T\left(
        \psi_\sigma({\bf p},t)\psi^{\dagger}_{\sigma'}({\bf p},0)
                \right)
    \right\rangle,
\end{equation}
where $\sigma,\sigma'$ are the non-relativistic, two-dimensional
nucleon spinor indices.
The relativistic and non-relativistic nucleon propagators can
then be related by substituting Eq.(\ref{N-psi}) into
Eq.(\ref{A2}):
\begin{equation}    \label{R-NR}
\frac1{2M_A}{\cal A}_{\alpha\beta}(p) =
U_{\alpha\sigma}\
        {\cal A}^{\rm NR}_{\sigma\sigma'}(\varepsilon,{\bf p})\
U^\dagger_{\sigma'\alpha'}\gamma^0_{\alpha'\beta} .
\end{equation}
Here the operator $U$ translates two-component spinors
into four-component spinors:
\begin{eqnarray}
\label{U}
U_{\alpha\sigma}\psi_\sigma
&=&
\left( \begin{array}{r}
       Z\,\psi \\
        \frac{
              {\mbox{\boldmath$\sigma\cdot$}\displaystyle\bf p} }
              {\displaystyle 2M}\,\psi
      \end{array}
\right)_{\!\alpha}.
\end{eqnarray}
Equation (\ref{R-NR}) can be used to
reduce the traces of the relativistic propagator ${\cal A}$
with the various Dirac structures in $\widehat{\cal G}^\beta$ to
expressions involving the non-relativistic propagator
${\cal A}_{\text{NR}}$:
\begin{mathletters}
\label{Trs}
\begin{eqnarray}
\frac1{2M_A}
\text{Tr} \left[ \gamma_0\gamma_5{\cal A}(p)
          \right]
&=&
\text{tr} \left[ \widehat{\cal S}_0\
                 {\cal A}_{\text{NR}}(\varepsilon,{\bf p})
          \right],\\
\frac1{2M_A}
\text{Tr} \left[ \gamma_j \gamma_5 {\cal A}(p)
          \right]
&=&
\text{tr} \left[ \widehat{\cal S}_j\
                 {\cal A}_{\text{NR}}(\varepsilon,{\bf p})
          \right], \\
\frac1{2M_A}
\text{Tr} \left[ i\gamma_5\sigma_{0 j}{\cal A}(p) \right]
&=&
\text{tr} \left[
          \left( - \sigma_j
             + \frac{\mbox{\boldmath$\sigma$}\cdot {\bf p}}{2M^2}
                   p_j
          \right) {\cal A}_{\text{NR}}(\varepsilon,{\bf p})
          \right], \\
\frac1{2M_A}
\text{Tr} \left[
        i\gamma_5\sigma_{i j}{\cal A}(p)\right]
&=&
\text{tr}\left[ \left( \frac{\sigma_i p_j-\sigma_j p_i}M
                \right) {\cal A}_{\text{NR}}(\varepsilon,{\bf p})
         \right],
\end{eqnarray}
\end{mathletters}%
where
\begin{mathletters}
\label{Shat}
\begin{eqnarray}
\widehat{\cal S}_0
&=& \frac{\mbox{\boldmath$\sigma$}{\bf\cdot p}}M,        \\
\widehat{\cal S}_j
&=&
\left( 1-\frac{{\bf p}^2}{2M^2} \right)
\sigma_j\
 +\ \frac{\mbox{\boldmath$\sigma$}{\bf\cdot p}}{2M^2}\ p_j,
\end{eqnarray}
\end{mathletters}%
and $i,j$ denote spatial indices.
The trace ``tr'' is taken with respect to the spin variable
in two-component space,
tr$\,{\cal O}\equiv {\cal O}_{\sigma\sigma}$.
All corrections to Eqs.(\ref{Trs}) are of order
$|{\bf p}|^3/M^3$ or higher.
One notices that the operators $\widehat{\cal S}_{0,j}$
have a structure similar
to the time and space components of the spin four-vector
$(0; {\mbox{\boldmath$\sigma$}})$ boosted
to a frame in which the nucleon has momentum ${\bf p}$.

{}From Eqs.(\ref{Trs}) and (\ref{Shat}) one can determine
the traces relevant for $g^A_{1,T}$ in Eqs.(\ref{g1TA}),
namely:
\begin{mathletters}
\label{TrG}
\begin{eqnarray}
& & \hspace*{-0.5cm}
\frac1{2M_A} \text{Tr}
\left[ {\cal A}^{\|}(p)\,
       \left(\widehat{\cal G}_0 + \widehat{\cal G}_z \right)
\right]                                         \nonumber\\
&=& \left( G^{(q)}
        + G^{(\gamma)}
        + \left( G^{(\sigma p)} - G^{(\sigma pq)} \right)
           \left(1 + \frac{p^2-M^2}{2M^2}\right)
    \right)
    \text{tr}
    \left[ {\cal A}^{\|}_{\text{NR}}(\varepsilon,{\bf p})\,
           \left( \widehat{\cal S}_0 + \widehat{\cal S}_z \right)
    \right], \\
& & \hspace*{-0.5cm}
\frac1{2M_A} \text{Tr}
\left[ {\cal A}^{\perp}(p)\,
       \widehat{\cal G}_\perp
\right]                                                 \nonumber\\
&=& \left( G^{(\gamma)}
        + G^{(\sigma p)} \left(1 + \frac{p^2-M^2}{2M^2}\right)
        + G^{(\sigma q)} \left(1 - \frac{p^2-M^2}{2M^2}\right)
    \right)
    \text{tr}
    \left[ {\cal A}^{\perp}_{\text{NR}}(\varepsilon,{\bf p})\,
           \widehat{\cal S}_\perp
    \right]                                             \nonumber\\
&+&
    \left(-G^{(q)}
        + G^{(\sigma q)} \left(1 - \frac{p^2-M^2}{2M^2}\right)
        + G^{(\sigma pq)} \left(1 + \frac{p^2-M^2}{2M^2}\right)
    \right)
    \text{tr}
    \left[ {\cal A}^{\perp}_{\text{NR}}(\varepsilon,{\bf p})\,
           \widehat{\cal T}_2
    \right],
\end{eqnarray}
\end{mathletters}%
where $\widehat{\cal S}_0 + \widehat{\cal S}_z$
and $\widehat{\cal S}_\perp$ are given by (\ref{Shat}),
and $p^2 \approx M^2+2M (\varepsilon-{\bf p}^2/2M)$ is the squared
nucleon four-momentum (the $\varepsilon^2$ term is dropped
since it introduces corrections of order ${\bf p}^4/M^4$).
The operator $\widehat{\cal T}_2$ is given by:
\begin{eqnarray}
\widehat{\cal T}_2
&=& - { {\bf p}_\perp \cdot {\bf S}^\perp \over M }
    \left(   \frac{ {\mbox{\boldmath$\sigma$}}{\bf \cdot p} }{M}
          + \sigma_z \left( 1 - \frac{p_z}{M} \right)
    \right),
\end{eqnarray}
where ${\bf p}_\perp$ is the transverse component of the nucleon
three-momentum vector:\ ${\bf p} = ({\bf p}_\perp, p_z)$,
and ${\bf S}^\perp$ defines the transverse spin quantization
axis, relative to the photon direction
(not to be confused with the spin operator $\widehat{\cal S}_\perp$).
Note that in the non-relativistic limit the structure
$G^{(p)}$ does not contribute to (\ref{TrG}).

An important observation which can be made from Eq.(\ref{TrG})
is that the nuclear structure functions are expressed in terms
of only \underline{two} combinations constructed from the $G^{(i)}$
in Eq.(\ref{DirSt}),
\begin{mathletters}
\label{g-off}
\begin{eqnarray}
g_1^N(x,p^2)
&=& G^{(q)}\
 +\ G^{(\gamma)}\
 +\ \left( G^{(\sigma p)} - G^{(\sigma pq)} \right)
    \left( 1 + \frac{p^2-M^2}{2M^2} \right),  \\
g_T^N(x,p^2)
&=& G^{(\gamma)}\
 +\ G^{(\sigma p)}
    \left(1 + \frac{p^2-M^2}{2M^2}\right)\
 +\ G^{(\sigma q)}
    \left(1 - \frac{p^2-M^2}{2M^2}\right),
\end{eqnarray}
so that also
\begin{eqnarray}
g_2^N(x,p^2)
&=& -G^{(q)}\
 +\ G^{(\sigma q)} \left(1 - \frac{p^2-M^2}{2M^2}\right)\
 +\ G^{(\sigma pq)} \left(1 + \frac{p^2-M^2}{2M^2}\right)
                                          \nonumber\\
&=& g_T^N(x,p^2)\ -\ g_1^N(x,p^2).
\end{eqnarray}
\end{mathletters}%
These can be considered as {\em definitions} of the
polarized nucleon structure functions in the off-mass-shell region
(c.f. Eqs.(\ref{g12on}) above) in the vicinity of $p^2 \approx M^2$.
Note that in the $p^2 = M^2$ limit they reduce directly to the
free nucleon structure functions defined in Eqs.(\ref{g12on}).

\subsection{Two-Dimensional Convolution}

The definitions of the off-shell structure functions in
Eqs.(\ref{g-off}) can now be utilized in deriving
convolution formulae for $g_{1,T}^A$.
After substituting Eqs.(\ref{TrG}) into (\ref{g1TA}),
we make use of the analytical properties of the nucleon
propagator for the integration over the momentum $p$.
Namely, we close the contour of integration in the
upper half of the complex $\varepsilon$ (or $p_0$) plane
and pick the poles of ${\cal A}_{\text{NR}}(\varepsilon)$
which correspond to the time ordering
$\theta(-t) \left\langle \psi^\dagger(0)\psi(t) \right\rangle$.
Finally, one obtains simplified versions of Eqs.(\ref{g1TA})
which relate the polarized nuclear structure functions directly
to those of nucleons:
\begin{mathletters}
\label{gNR}
\begin{eqnarray}
xg_1^A(x)
&=& \int\!\frac{d\varepsilon d^3{\bf p}}{(2\pi)^4}\
    {\rm tr} \left[ {\cal P}^{\|}(\varepsilon,{\bf p})\,
                    \left( \widehat{\cal S}_0 + \widehat{\cal S}_z
                    \right)
             \right] x'g_1^N(x',p^2)\ ,         \label{gNR1}\\
xg_T^A(x)
&=& \int\!\frac{d\varepsilon d^3{\bf p}}{(2\pi)^4}\
    \left(
    {\rm tr} \left[ {\cal P}^{\perp}(\varepsilon,{\bf p})\,
                    \widehat{\cal S}_\perp
             \right] x'g_T^N(x',p^2)\
 +\ {\rm tr} \left[ {\cal P}^{\perp}(\varepsilon,{\bf p})\,
                    \widehat{\cal T}_2
             \right] x'g_2^N(x',p^2)
    \right),                                   \label{gNRT}
\end{eqnarray}
\end{mathletters}%
where here the nuclear structure functions are expressed as
functions of the standard Bjorken variable
$x=Q^2/2Mq_0=x_A M_A/M$.
For a general polarization state, the nuclear spectral function,
${\cal P}(\varepsilon,{\bf p})$, is defined by:
\begin{equation}\label{S}
{\cal P}_{\sigma\sigma'}(\varepsilon,{\bf p})
= \sum_n
  \psi_{n,\sigma}({\bf p}) \psi_{n,\sigma'}^*({\bf p})\, \
  2\pi\delta
  \left( \varepsilon - E_0(A) + E_n(A-1,-{\bf p} ) \right),
\end{equation}
where the summation is performed over the complete set of
states with $A-1$ nucleons.
The functions
$\psi_{n,\sigma}({\bf p})
= \left\langle (A-1)_n,-{\bf p}|\psi_\sigma(0)|A \right\rangle$
give the probability amplitude to find in the nuclear ground state
a nucleon with polarization $\sigma$ and the remaining $A-1$
nucleons in a state with total momentum $-{\bf p}$
($n$ labels all other quantum numbers).
The non-relativistic energies of the target ground state
and the $A-1$ residual nucleons are denoted
$E_0(A)$ and $E_n(A-1)$, respectively.
Summing over polarizations $\sigma$ and $\sigma'$,
the spectral function ${\cal P}(\varepsilon,{\bf p})$ is normalized
to the number of nucleons $A$:
\begin{eqnarray}
\int\!\frac{d\varepsilon d^3{\bf p}}{(2\pi)^4}\
{\rm tr} \left[ {\cal P}(\varepsilon,{\bf p}) \right]
&=& A.
\end{eqnarray}

Equations (\ref{gNR}) can be written in a more familiar
form as two-dimensional convolutions of the off-shell nucleon
structure functions in Eqs.(\ref{g-off}) and nucleon momentum
distribution functions $D(y,p^2)$:
\begin{mathletters}
\label{convol}
\begin{eqnarray}
g_1^A(x)
&=& \int\!dp^2\int\limits_x\frac{dy}y\,
    D_1(y,p^2)\,g_1^N\!\left(\frac{x}y,p^2\right), \label{con1}\\
g_T^A(x)
&=& \int\!dp^2\int\limits_x\frac{dy}y\,
    \left[ D_T(y,p^2)\,g_T^N\!\left(\frac{x}y,p^2\right)\
               +\ D_{T2}(y,p^2)\,g_2^N\!\left(\frac xy,p^2\right)
           \right],                           \label{conT}
\end{eqnarray}
\end{mathletters}%
where $y = (p_0 + p_z) / M$ is the fraction of the
light-cone momentum of the nucleus carried by the
interacting nucleon.
The nucleon distribution functions are given by:
\begin{mathletters}
\label{D}
\begin{eqnarray}
D_1(y,p^2)
&=& \int\!\frac{d\varepsilon d^3{\bf p}}{(2\pi)^4}\
    {\rm tr}
    \left[ {\cal P}^{\|}(\varepsilon,{\bf p})
           \left( \widehat{\cal S}_0 + \widehat{\cal S}_z \right)
    \right]
    \delta \left( y - \frac{M+\varepsilon+p_z}{M} \right)
    \delta \left( p^2 - \mu^2 \right),    \\
D_T(y,p^2)
&=& \int\!\frac{d\varepsilon d^3{\bf p}}{(2\pi)^4}\
    {\rm tr}
    \left[ {\cal P}^{\perp}(\varepsilon,{\bf p})\
           \widehat{\cal S}_\perp
    \right]
    \delta \left( y - \frac{M+\varepsilon+p_z}{M} \right)
    \delta \left( p^2 - \mu^2 \right),    \\
D_{T2}(y,p^2)
&=& \int\!\frac{d\varepsilon d^3{\bf p}}{(2\pi)^4}\
    {\rm tr}
    \left[ {\cal P}^{\perp}(\varepsilon,{\bf p})\
           \widehat{\cal T}_2
    \right]
    \delta \left( y - \frac{M+\varepsilon+p_z}{M} \right)
    \delta \left( p^2 - \mu^2 \right),
\end{eqnarray}
\end{mathletters}%
where $\mu^2 \equiv M^2+2M(\varepsilon-{\bf p}^2/2M)$.
We observe that $g_1^A$ is expressed entirely in terms of $g_1^N$,
while $g_T^A$ receives contributions from $g_T^N$
as well as from $g_2^N$.
Consequently, the $g_2^A$ structure function will
receive contributions from $g_1^N$ in addition to
$g_2^N$.

\section{Expansion Formulae}
\label{expansion}

The behavior of the nucleon distribution functions in Eqs.(\ref{D})
is governed by the nuclear spectral function
${\cal P}(\varepsilon,{\bf p})$.
The region of importance for ${\cal P}(\varepsilon,{\bf p})$ is
$|{\bf p}|\lesssim p_F,\ |\varepsilon|\lesssim p_F^2/2M$,
where $p_F$ is a characteristic momentum which determines
the momentum distribution of nucleons in the nucleus.
For heavy nuclei this is the Fermi--momentum, $p_F\approx 300\,$MeV.
For a light nucleus, such as the deuteron, the analogous parameter
can be determined from the average kinetic energy $T$ as
$\sqrt{M\,T}\approx 140\,$MeV.
Therefore the nucleon distribution functions (\ref{D}) are strongly
peaked about the light-cone momentum fraction $y=1$
and the on-mass-shell point $p^2 = M^2$.

This property of the distribution functions allows us to obtain
approximate expressions for the nuclear structure functions in
Eqs.(\ref{convol}).
The $p^2$ dependence of the off-shell structure functions
can be first approximated by expanding $g_{1,T}^N(x/y,p^2)$
in a Taylor series around $p^2 = M^2$:
\begin{eqnarray}
g_{1,T}^N\left({x\over y},p^2\right)
&\approx& g_{1,T}^N\left({x\over y}\right)\
 +\ (p^2 - M^2)
    \left.{ \partial g_{1,T}^N(x/y,p^2) \over \partial p^2 }
    \right|_{p^2=M^2},
\label{g1Tp2}
\end{eqnarray}
where $g_{1,T}^N(x/y) \equiv g_{1,T}^N(x/y,p^2=M^2)$
is the structure function of the (physical) on-mass-shell nucleon
(see Eqs.(\ref{g12on}) and (\ref{g-off})).
Expanding $g_{1,T}^N(x/y) / y$ in Eqs.(\ref{convol}) around $y=1$,
integrating the result term by term, and keeping terms
up to order $\varepsilon/M$ and ${\bf p}^2/M^2$,
we obtain simple expansion formulae similar to those
used in the analysis of unpolarized
nuclear structure functions \cite{AKV,KUL,KPW}.
For the structure function $g_1^A$ we find:
\begin{eqnarray}
\label{exp1}
\frac1A g_1^A(x)
&\approx& C_1^{(0)}\, g_1^N(x)
 + C_1^{(1)} \left(xg_1^N(x)\right)'
 + C_1^{(2)} x\left(xg_1^N(x)\right)''
 + C_1^{(3)} \left.\frac{\partial g_1^N(x,p^2)}{\partial\ln p^2}
                \right|_{p^2=M^2},
\end{eqnarray}
where the derivatives are taken with respect to $x$,
and the coefficients are:
\begin{mathletters}
\label{C1}
\begin{eqnarray}
C_1^{(0)}
&=& \left\langle \widehat{\cal S}_0 + \widehat{\cal S}_z
    \right\rangle^{\|}, \\
C_1^{(1)}
&=& \left\langle
    \left( \widehat{\cal S}_0 + \widehat{\cal S}_z \right)
    \left( \frac{p_z^2}{M^2} - \frac{p_z+\varepsilon}M \right)
    \right\rangle^{\|},                                  \\
C_1^{(2)}
&=& \left\langle
    \left( \widehat{\cal S}_0 + \widehat{\cal S}_z \right)
    \frac{p_z^2}{2M^2}
    \right\rangle^{\|},                                   \\
C_1^{(3)}
&=& \left\langle
    \left( \widehat{\cal S}_0 + \widehat{\cal S}_z \right)
    \frac2M \left(\varepsilon-\frac{{\bf p}^2}{2M}\right)
    \right\rangle^{\|},
\end{eqnarray}
\end{mathletters}%
with $\widehat{\cal S}_0 + \widehat{\cal S}_z$ obtained from
Eqs.(\ref{Shat}).
Here the averaging $\langle \cdots \rangle$ denotes:
\begin{eqnarray}
\left\langle {\cal O} \right\rangle^{\|}
&\equiv& \frac1A\int\!\frac{d\varepsilon d^3{\bf p}}{(2\pi)^4}\
         {\rm tr}
         \left[ {\cal P}^{\|}(\varepsilon,{\bf p})\, {\cal O}
         \right].
\end{eqnarray}
Analogously, the structure function $g_T^A$ is given by:
\begin{eqnarray}
\label{expT}
\frac1A g_T^A(x)
&\approx& C_T^{(0)}\, g_T^N(x)\
 +\ C_T^{(1)}\, \left(xg_T^N(x)\right)'\
 +\ C_T^{(2)}\, x\left(xg_T^N(x)\right)''\
 +\ C_T^{(3)}\, \left.\frac{\partial g_T^N(x,p^2)}{\partial\ln p^2}
                \right|_{p^2=M^2}                       \nonumber\\
&+& C_{T2}^{(0)}\, g_2^N(x)\
 +\ C_{T2}^{(1)}\, \left(xg_2^N(x)\right)'.
\end{eqnarray}
The coefficients $C_T^{(i)}$ are identical to $C_1^{(i)}$
except for the replacements
$\left\langle {\cal O} \right\rangle^{\|} \rightarrow
 \left\langle {\cal O} \right\rangle^{\perp}$
and
$\widehat{\cal S}_0 + \widehat{\cal S}_z
 \rightarrow \widehat{\cal S}_\perp$,
and the coefficients $C_{T2}^{(i)}$ are:
\begin{mathletters}
\label{CT2}
\begin{eqnarray}
C_{T2}^{(0)}
&=& \left\langle \widehat{\cal T}_2
    \right\rangle^{\perp},      \\
C_{T2}^{(1)}
&=& - \left\langle \widehat{\cal T}_2\ {p_z \over M}
      \right\rangle^{\perp}.
\end{eqnarray}
\end{mathletters}%
We stress that the matrix elements $C_1^{(i)}$ are calculated
for longitudinally polarized targets, while for
$C_T^{(i)}$ and $C_{T2}^{(i)}$ the target is polarized
in a direction transverse to the photon direction.
Once $g_1^A$ and $g_T^A$ are calculated, the difference
can be taken and the approximate result for $g_2^A$ obtained.
Corrections to $g^A_{1,2}$ evaluated via Eqs.(\ref{exp1}) and
(\ref{expT}) are of order ${\bf p}^3/M^3$ or higher.

In the derivation of Eqs.(\ref{exp1}) and (\ref{expT})
one neglects the lower limit in the $y$-integration
in Eqs.(\ref{convol}), namely the condition $x/y\le 1$.
{}From Eqs.(\ref{D}) one can easily see that this condition
gives practically no restriction on the integration region
in (\ref{C1},\ref{CT2}) for $1-x > p_F/M$.
For heavy nuclei, the expansion formulae can be used safely
up to $x \approx 0.7$.
For light nuclei such as the deuteron, which we consider
in the next section, Eqs.(\ref{exp1}) and (\ref{expT})
are expected to be reliable for $x \lesssim 0.8$.

\section{The Deuteron}
\label{deuteron}

In this section we consider the application of the results
of Sections \ref{nr-limit} and \ref{expansion} to the case
of a deuterium nucleus.
As mentioned in Section~\ref{introdu},
deep-inelastic scattering from
polarized deuterons is an important means of obtaining
information on the polarized neutron structure functions
$g_{1,2}^n$.
Extracting information on $g_{1,2}^n$ from deuterium data
is only meaningful, however, if one has a reliable method
of subtracting the relevant nuclear effects.

The several previous attempts to account for nuclear
effects in the deuteron (for the $g_1^D$ structure function)
have not always yielded consistent results.
Some early attempts \cite{FS} were made within a time-ordered
framework in the infinite momentum frame, which unfortunately
in practice was problematic due to the lack of knowledge about
deuteron wavefunctions in this frame.
Subsequent analyses \cite{WOL} utilized convolution formulae
obtained in direct analogy with the convolution model for
unpolarized scattering.
Namely, a one-dimensional convolution formula was used
with the same non-relativistic ``flux factor'',
$(1+p_z/M)$, as appears in the unpolarized deuteron
$F_{2D}$ structure function \cite{KUL,KPW,JM,BT}.
Other attempts \cite{KAP2} were based on an operator product
expansion at the nucleon level \cite{KAP3}, however these led
to different operators to those in Eq.(\ref{gNR1}).
It is important, therefore, to clarify which operators,
or ``flux factors'', are relevant for $g_1^D$ when all
terms up to order ${\bf p}^2/M^2$ are consistently kept.

In addition to $g_1^D$, there is considerable practical
value in understanding the nuclear effects on $g_T^D$
(and hence $g_2^D$),
to which little attention has been paid thus far.
In view of upcoming experiments which will for the first time
measure the deuteron $g_2^D$ structure function \cite{PETR},
an estimate of the relevant nuclear corrections to
$g_T^D$ is urgently needed.

As a guide to evaluating the most efficient method
for the nuclear data analysis, a comparison of the
results for $g_{1,2}^D$ calculated using the convolution
(\ref{convol}) and expansion (\ref{exp1},\ref{expT})
formulae will indicate the reliability of the latter
approach to the deuteron, which to date has not been
explicitly tested.
Essentially the only ingredient needed to evaluate
the coefficients in Eqs.(\ref{exp1}) and (\ref{expT}),
as well as the traces in the distribution functions
of Eqs.(\ref{D}), is the deuteron wavefunction.
Before discussing the details of the deuteron case, however,
we must first fix the nucleon inputs that will be used in the
subsequent numerical calculations.

\subsection{Nucleon Structure Function Input}
\label{sf-input}

For the structure functions and their derivatives
we shall rely on experimental results where appropriate,
and use model input where data are not yet available.
For the proton and neutron $g_1$ structure functions we
use the recent parametrization from Ref.\cite{GS} of the
SLAC \cite{SLAC}, EMC \cite{EMC} and SMC \cite{SMCP} proton,
the SLAC-E142 neutron (Helium-3) \cite{E142},
and SMC deuteron \cite{SMCD} data.
As an illustration of the quality of the fit,
we plot in Fig.\ref{F_NSF}(a) the $xg_1^{p,n}$ structure
functions at $Q^2 = 4$ GeV$^2$, compared with the data.

For the $g_2$ structure function the study of nuclear effects
is more problematic, since this receives contributions from
both twist-2 and 3 operators, the latter of which contain
quark-gluon interactions and also explicitly depend on
quark masses.
Following the standard decomposition of $g_2^N$ into the
different twist components, we write:
\begin{eqnarray}
g_2^N(x) &=& g^{N(WW)}_2(x)\ +\ \overline{g}^N_2(x),  \label{g2N}
\end{eqnarray}
where the twist-2 part is given by the Wandura-Wilczek relation
\cite{WW}:
\begin{eqnarray}
g^{N(WW)}_2(x)
&=& \int_x^1 \frac{dy}{y}
    \left( 1 - \delta(1 - x/y) \right)\
    g_1^N(y),                                            \label{WW}
\end{eqnarray}
and satisfies the Burkhardt--Cottingham sum rule \cite{BC}:
\begin{eqnarray}
\int_0^1 dx\ g_2^N(x) &=& 0.            \label{BC}
\end{eqnarray}

The twist-3 piece ($\overline{g}_2^N$) of $g_2^N$ is at present
not very well determined at all.
There is disagreement even about the magnitude and sign
of its moments \cite{JJI,QCDSUM}.
Based on a covariant parton model approach,
Jackson, Roberts and Ross \cite{JRR} argued in favor of
a very small twist-3 component.
Within bag models, on the other hand, one finds \cite{JJI,STRAT}
quite a sizable $\overline{g}_2^N$ contribution
compared with the Wandura-Wilczek term.
An additional problem is how to relate the structure function
calculated in the bag (or some other) model, which one assumes
to be applicable at some low resolution scale
$Q^2 = {\cal O}(\Lambda_{QCD}^2)$, to that
appropriate to DIS experiments ($Q^2 \agt 5$ GeV$^2$).
One prescription \cite{GRV} is to simply assume the
validity of the Altarelli-Parisi evolution equations
down to very low $Q^2$.
Even within this pragmatic approach,
while the evolution of the twist-2 component follows that of
$g_1$, which is understood, for the $\overline{g}_2^N$ piece
only the $N=$ 2 and 4 moments can be handled exactly.
The solution adopted in \cite{STRAT} was to use an approximate
solution of the evolution equations that was derived in the
large-$N_C$ limit.

Since the only available data \cite{SMC_g2p} on the $g_2^p$
structure function cannot yet unambiguously discriminate
between the various models of $\overline{g}_2$, we will estimate
the size of the nuclear effects for several models.
To cover the potential range of results for
$\overline{g}_2^N$, we take $\overline{g}^N_2=0$, as suggested
by Ref.\cite{JRR}, and also the bag model predictions,
evolved to $Q^2 \sim 5$ GeV$^2$ using the prescription adopted
in Ref.\cite{STRAT}.
As an indication of the possible variation of the total $g_2^N$
structure function with different twist-3 components,
we plot in Fig.\ref{F_NSF}(b) $g_2^N$ for these two cases.
Also shown is the sum, $g_T^N = g_1^N + g_2^N$, which will
be relevant in the actual evaluation of the nuclear structure
function, Eqs.(\ref{convol},\ref{exp1},\ref{expT}).
One can see that while the effect of the twist-3 component
is certainly not negligible, it does not alter drastically
the overall shape and sign of $g_2^N$ and $g_T^N$.

{}From the two-dimensional convolution equations
(\ref{convol}) it is clear that a consistent
description of nuclear effects to order ${\bf p}^2/M^2$
requires modeling in addition the $p^2$ dependence
of the off-shell nucleon structure functions
$g_{1,2}^N(x,p^2)$.
To this order of accuracy this can be achieved by determining
the slope with respect to $p^2$ at the on-mass-shell point,
Eq.(\ref{g1Tp2}).
One could, for example, formulate $g^N_{1,2}$
in terms of relativistic quark--nucleon vertex functions
as described in Refs.\cite{MPT,MST,MSTD,MM,MW}, and calculate
the derivative directly.
Alternatively, to obtain a quick estimate of the overall order
of magnitude of the off-shell effect, we can extend the model of
Ref.\cite{KPW}, which is based on a dispersion representation of
the unpolarized nucleon structure function, to the polarized case.

Within the latter approach, in the impulse approximation
for the quarks the $g_1^N$ structure function can be written:
\begin{eqnarray}
g^N_{1}(x,p^2)
&=& \int ds\ \int_{-\infty}^{k^2_{max}(x,p^2)} dk^2\
    \rho(k^2,s,p^2,x),                          \label{g1p2}
\end{eqnarray}
where $k^2_{max}(x,p^2) = -x s / (1-x) + x p^2$ is the
kinematical maximum of the quark momentum squared $k^2$,
$s = (p-k)^2$ is the center-of-mass energy squared of the
``spectator'' quark system,
and $\rho$ is the quark spectral function extended
to the nucleon off-mass-shell region.
Following Ref.\cite{KPW} we assume that $\rho$ has no explicit
$x$-dependence, and that the spectrum in $s$ can be approximated
by that calculated for a single effective mass $\bar s$,\
$\rho \propto \delta\left( s - \bar s \right)$.
At moderate $Q^2 \sim 5-10$ GeV$^2$ one finds typically
$\bar s \simeq 2$ GeV$^2$ \cite{KPW}.
Taking a factorized $p^2$ and $k^2$ dependence
in $\rho$\ then gives $g_1^N$ as a product of two functions:
\begin{eqnarray}
g^N_{1}(x,p^2)
&=& \varphi(p^2)\ {\cal F} \left( k^2_{max}(x,p^2) \right),
\label{funnyg1}
\end{eqnarray}
normalized such that $\varphi(M^2) = 1$ and hence
${\cal F} (k^2_{max}(x,M^2)) = g_1^N(x)$.
The explicit $p^2$-dependence in the function $\varphi(p^2)$
is dynamical in origin, while the $p^2$-dependence
in the function ${\cal F}(k^2_{max})$ that enters through
the upper limit of the $k^2$ integration
is purely kinematical.
After some algebra, the derivative of $g^N_{1}(x,p^2)$
with respect of $p^2$ can then be written:
\begin{eqnarray}
\left. { \partial g^N_{1}(x,p^2) \over \partial p^2 }
\right|_{p^2=M^2}
&=& g^N_1(x)\
    \left. {\partial \varphi(p^2) \over \partial p^2 }
    \right|_{p^2=M^2}\
 +\ \left( g^N_{1}(x) \right)'\
    { x\ (1-x)^2 \over M^2\ (1-x)^2\ -\ \bar s }\ .
\label{dg1dp2}
\end{eqnarray}

To determine the slope of the $\varphi(p^2)$ at $p^2=M^2$
we assume that the first moment of the non-singlet part
of $g_1^N$ is not renormalized off-shell
(at least in the vicinity of $p^2 \approx M^2$):
\begin{eqnarray}
\int_0^1 dx\ \left. { \partial g_1^N(x,p^2) \over \partial p^2 }
             \right|_{p^2=M^2}
&=& 0.
\label{p2norm}
\end{eqnarray}
The motivation for this condition is that since the
axial U(1) anomaly \cite{U1} is absent in the non-singlet
sector, in the chiral limit the axial charge of the nucleon
is a conserved quantity.
For the singlet component of $g_1^N$ little is known about
how the effects of the axial anomaly extrapolate into
the off-shell region, although one would not expect dramatic
consequences as long as $p^2 \approx M^2$.
To satisfy Eq.(\ref{p2norm}) with $\bar s = 2$ GeV$^2$
one needs a slope of approximately
$\left. \partial \varphi(p^2) / \partial p^2
 \right|_{p^2=M^2}
 \approx -0.16$  GeV$^{-2}$.
For larger values of $\bar s$, $\bar s \simeq 3$ GeV$^2$,
the slope becomes $\simeq -0.1$ GeV$^{-2}$.

The extension of the above off-shell model to the $g_2^N$
structure function is more problematic.
For example, there is no known justification for a
normalization condition such as in Eq.(\ref{p2norm})
to be valid for $g_2^N$.
Furthermore, there is little knowledge about how the
higher twist correlation, or final state interaction,
effects involving the (nucleon--quark) ``spectator''
system would modify the effective mass $\bar s$.
For these reasons we postpone for the time being a more
detailed discussion about the off-shell dependence in $g_2^N$.

With these inputs for the nucleon structure function
we can now proceed to evaluate numerically the structure
functions of the deuteron.

\subsection{Convolution Results}

For the deuteron, the traces in Eqs.(\ref{D}) can be
expressed in terms of the deuteron wavefunctions
$\Psi_{m=+1}({\bf p})$ with spin projection $m=+1$
along the axis of quantization (see Appendix~\ref{DI})
\footnote{Note that the argument in the deuteron wavefunction
is the relative nucleon momentum, which in the deuteron rest
frame coincides with the single particle momentum ${\bf p}$.}:
\begin{eqnarray}
\label{SOD}
\text{tr\,}\left[{\cal P}(\varepsilon,{\bf p})\ {\cal O}\right]_D
&=&
\Psi^\dagger_{+1}({\bf p})
\left( {\cal O}^{(p)} + {\cal O}^{(n)}
\right)
\Psi_{+1}({\bf p})\ \
2\pi\ \delta(\varepsilon - \epsilon_D + {\bf p}^2/2M),
\end{eqnarray}
where $\epsilon_D$ is the deuteron binding energy.
The expectation values in Eq.(\ref{SOD}) with the operators
${\cal O} = \widehat{\cal S}_0 + \widehat{\cal S}_z,\
            \widehat{\cal S}_\perp$ and
           $\widehat{\cal T}_2$
determine the deuteron distribution functions.
{}From Eq.(\ref{genME})
(for the axis of quantization parallel to the $z$-direction)
we find:
\begin{mathletters}
\label{DME}
\begin{eqnarray}
\Psi^{(\|)\dagger}_{+1}({\bf p})
\left( \widehat{\cal S}_0 + \widehat{\cal S}_z \right)
\Psi^{(\|)}_{+1}({\bf p})\,
&=&
\frac{4\pi^2}{{\bf p}^2}
\left\{
\left( 1-\frac{{\bf p}^2}{2M^2} \right)
\left[ u^2 + \frac{uw}{\sqrt2} \left( 1-3\widehat{p}_z^2 \right)
     + w^2 \left( \frac32\widehat{p}_z^2 - 1 \right)
\right]
\right.         \nonumber\\
&& \hspace*{1cm} +\
\left.
{ p_z \over M } \left( 1 + \frac{p_z}{2M}\right)
\left[ u-\frac w{\sqrt2} \right]^2
\right\},                                       \label{DME1}
\end{eqnarray}
where $u(p)$ and $w(p)$ are the $S$- and $D$-state wavefunctions
in momentum space, and $p_z = |{\bf p}| \widehat{p}_z$, with
$\widehat{p}_z~=~\cos\theta$ ($\theta$ is the angle between
$\widehat{\bf p}$ and the $z$-axis).

For a deuteron polarized in a direction transverse
relative to the photon direction, $S~=~S^\perp$, we have:
\begin{eqnarray}
\Psi^{(\perp)\dagger}_{+1}({\bf p})\,
\widehat{\cal S}_\perp\,
\Psi^{(\perp)}_{+1}({\bf p})\,
&=&
\frac{4\pi^2}{{\bf p}^2}
\left\{
\left( 1-\frac{{\bf p}^2}{2M^2} \right)
\left[
  u^2
+ \frac{uw}{\sqrt2}
  \left( 1-3 (\widehat{{\bf p}}_\perp \cdot {\bf S}^\perp)^2 \right)
\right.
\right.        \nonumber\\
& & 
\left.
\left.
+ w^2
  \left( \frac32 (\widehat{{\bf p}}_\perp \cdot {\bf S}^\perp)^2 - 1
  \right)
\right]
+
\frac{ \left( {\bf p}_\perp \cdot {\bf S}^\perp \right)^2 }{ 2 M^2 }
\left[ u-\frac w{\sqrt2} \right]^2
\right\},               \label{DME2}\\
\Psi^{(\perp)\dagger}_{+1}({\bf p})\,
\widehat{\cal T}_2\,
\Psi^{(\perp)}_{+1}({\bf p})\,
&=&
- \frac{4\pi^2}{|{\bf p}|}
{ \left( \widehat{{\bf p}}_\perp \cdot {\bf S}^\perp \right)^2
  \over M }
\left\{ {3\over2} \left( 1 - {p_z \over M} \right) \widehat{p}_z
        w \left( w - \sqrt{2} u \right)\
\right.                                                 \nonumber\\
& & \hspace*{3.5cm} +\
\left.
{ |{\bf p}| \over M } \left( u - {w \over \sqrt{2}} \right)^2
\right\}.               \label{DME3}
\end{eqnarray}
\end{mathletters}%

Substituting these results into Eqs.(\ref{D}), we can now
evaluate the $g_{1,T}^D$ structure functions of the deuteron.
In Fig.\ref{F_C1} we plot the ratio $g_1^D / g_1^N$ as a
function of $x$, for $Q^2 = 4$ GeV$^2$, using wavefunctions from
several different models of deuteron structure
\cite{PARIS,BONN,BG}.
The differences between the curves at small $x$ are
due mainly to the different deuteron $D$-state
probabilities, namely
5.8\% for the Paris,
4.3\% for the Bonn (full model) and
4.7\% for the Buck/Gross (pseudo-vector coupling) models.
In calculating the curves in Fig.\ref{F_C1} we have made the
approximation $g_1^N(x,p^2) \approx g_1^N(x,M^2)$.
Within the model of Ref.\cite{MPT} it was found that this
approximation is good to within 0.5\% for $x \alt 0.8$.
We shall illustrate the effects of the
$\partial g_1^N/\partial p^2$ term
evaluated using the off-shell model of Section~\ref{sf-input}
in the next section.

The transverse structure function ratio, $g_T^D/g_T^N$,
is shown in Fig.\ref{F_CT}, at $Q^2 = 4$ GeV$^2$,
calculated for the Paris wavefunction \cite{PARIS} (solid curve)
\footnote{Note that plotting the ratio of $g_2$ structure
functions is not very instructive since $g_2^N$ changes sign
--- to obtain the nuclear effects on $g_2$ alone one can
subtract the $g_1$ component determined above.}.
The result turns out to be quite similar to the ratio of the
$g_1$ structure functions in Fig.\ref{F_C1} (dashed curve),
the main difference being at large $x$, where the $g_T$
ratio rises above unity earlier.
The faster rise is even more pronounced for
the ratio of the twist-2 components of $g_T$ (dotted curve).
At intermediate $x$ values ($x \alt 0.5$) the $g_T^D/g_T^N$
ratio appears mostly independent of the model for $g_2^N$,
so that one may reasonably safely extract information on
$g_T^N$ (and hence $g_2^N$) from the transverse deuteron data.

\subsection{Expansion Results}

In the range of $x$ at which most of the data are taken,
namely $x \alt 0.6$, the arguments given in Section~\ref{expansion}
would suggest that the expansion formulae in Eqs.(\ref{exp1})
and (\ref{expT}) should be excellent approximations to the
full convolution results.
Since these would simplify the analysis of the deuterium data,
one may then take advantage of the simple expansion
approximations in this region of $x$.

To test the reliability of the expansion approach, we must
calculate the coefficients $C_1^{(i)}$, $C_T^{(i)}$ and
$C_{T2}^{(i)}$ for a deuteron target.
Using the matrix element in Eq.(\ref{DME1}) we obtain
explicit expressions for the coefficients in
Eqs.(\ref{C1}) of the various terms in the
structure function $g_1^D(x)$:
\begin{mathletters}
\begin{eqnarray}\label{C1D}
C_1^{(0)}
&=& 1-\frac32 P_D\
 -\ \frac2{3M}\left(T_0+\frac1{\sqrt2}T_{02}-T_2\right),\\
C_1^{(1)}
&=& \left(1-\frac32 P_D\right)\frac{|\epsilon_D|}M\
 +\ \frac1M\left(T_0+\frac{2\sqrt2}5 T_{02}-\frac9{10}T_2\right),\\
C_1^{(2)}
&=& \frac1{3M}
    \left(T_0-\frac{2\sqrt2}5 T_{02}-\frac1{10}T_2\right),\\
C_1^{(3)}
&=& -\frac2M
\left[ \left( 1-\frac32 P_D \right) |\epsilon_D| + 2T_0-T_2
\right],
\end{eqnarray}
\end{mathletters}%
where $P_D=\int dp\,w^2(p)$ is
the $D$-state probability in the deuteron,
with $p \equiv |{\bf p}|$ (not to be confused with
the momentum four-vector $p$ used above).
Here $T_0$, $T_2$ and $T_{02}$ represent the
average nucleon kinetic energies associated with each component
of the deuteron wavefunction:
\begin{mathletters}
\begin{eqnarray}
T_0    &=& \int_0^{\infty} dp\,\frac{{\bf p}^2}{2M}\ u^2(p),\\
T_2    &=& \int_0^{\infty} dp\,\frac{{\bf p}^2}{2M}\ w^2(p),\\
T_{02} &=& \int_0^{\infty} dp\,\frac{{\bf p}^2}{2M}\ u(p)\ w(p).
\end{eqnarray}
\end{mathletters}%

To illustrate the role of the various terms
in the expansion we plot in Fig.\ref{F_TERMS1} the
zeroth order contribution, proportional to $C_1^{(0)}$,
together with the higher order terms, scaled by
a factor 100.
Evident at large values of $x$ ($x \sim 0.8$)
is the role of the second derivative term,
proportional to $C_1^{(2)}$, which gives
the characteristic rise (due to Fermi motion)
of the structure function ratio
$g_1^D/g_1^N$ as $x \rightarrow 1$,
see Fig.\ref{F_E1}.
The trough in the ratio at $x \sim 0.6$ arises from
the single differential term proportional
to $C_1^{(1)}$ which is large and negative in
this region.
Although it does not give rise to any uniquely distinct
features in the structure function ratio, it is
clear that the off-shell component (dotted curve in
Fig.\ref{F_TERMS1}) is of the same order of magnitude
as the other higher order corrections,
and must be included in any precision analysis
of nuclear effects in the deuteron.

In Fig.\ref{F_E1} we compare the performance of
the expansion formula for the ratio $g_1^D/g_1^N$
with several other methods of computation,
neglecting for the moment the off-shell contributions.
The simplest approach (and the one used by the
SMC in their analysis \cite{SMCD}) is to use
a constant depolarization factor, $(1 - 3/2\,P_D)$,
which roughly corresponds to the first term in
the Taylor series in Eq.(\ref{exp1}).
In fact, the order ${\bf p}^2/M^2$ correction to
$(1 - 3/2\,P_D)$ is of the order of 10\%,
i.e. an overall correction to the structure function
ratio of $\sim 0.5\%$.
At present this is still smaller than the uncertainty
in $P_D$ between the different models.
Compared with the convolution results, one sees
that the expansion formula works remarkably well
for $x \alt 0.7$, where the two results are almost
indistinguishable.
As expected, for $x \agt 0.7$ the expansion curve
overshoots the convolution result, which is understood
from the fact that in the expansion formula one is
neglecting the lower limit of integration, namely $x$,
of $D(y)$ over $y$, and replacing it by zero.

To examine the effect of the off-shell $C_1^{(3)}$
term in Eq.(\ref{exp1}) on the $g_1$ ratio,
we plot in Fig.\ref{F_E1OFF} the ratio with
$\partial g_1^N / \partial p^2 = 0$ (dashed)
and with the slope determined through Eq.(\ref{dg1dp2})
(solid) with the mass parameter $\bar s = 2$ GeV$^2$.
The overall effect on the shape of the ratio is quite
minimal, and the trend follows the shape of the off-shell
component in Fig.\ref{F_TERMS1}.
Note that within this model the off-shell curve at
large $x$ ($x \agt 0.7$) approaches the on-shell limit,
although in this region the expansion approximation itself
is no longer accurate.
Within the relativistic model of Ref.\cite{MPT} (dotted curve),
the magnitude of the (negative) off-shell effects was found
to increase rapidly beyond $x \sim 0.8$, which is a further
reason why the expansion curves tend to be larger
at very large $x$.
In the intermediate-$x$ region, on the other hand,
there is little to distinguish all of the curves
for $0.2 \alt x \alt 0.7$.

For the transversely polarized function $g_T^D(x)$,
using Eqs.(\ref{DME2},\ref{DME3}) we obtain:
\begin{mathletters}
\begin{eqnarray}
C_T^{(0)}
&=& C_1^{(0)},                                                  \\
C_T^{(1)}
&=& \left( 1 - \frac32 P_D \right) \frac{|\epsilon_D|}M
 +  \frac{5}{3M}
    \left( T_0 + {2\sqrt2\over 25} T_{02} - {29\over 50}T_2
    \right),                                                    \\
C_T^{(2)}
&=& {1\over 3M}
    \left( T_0 + {\sqrt2 \over 5} T_{02} - {7\over 10} T_2
    \right),                                                    \\
C_T^{(3)}
&=& C_1^{(3)},                                                  \\
C_{T2}^{(0)}
&=& -{2\over 3M}
     \left( T_0 - {7\sqrt{2}\over 10} T_{02} + {1\over5} T_2
     \right),                                                   \\
C_{T2}^{(1)}
&=& {1\over 5M} \left( T_2 - \sqrt2 T_{02} \right).
\end{eqnarray}
\end{mathletters}%
The various components of $xg_T^D$ in Eq.(\ref{expT})
are shown in Fig.\ref{F_TERMST}.
As in Fig.\ref{F_E1}, the higher order corrections
are scaled by 100.
(As discussed in Section~\ref{sf-input},
we do not consider the off-shell
$C_T^{(3)}$ term here.)
While small, the corrections proportional to $xg_2^N$ and
its derivative are still of the same order of magnitude
as the $g_T^N$-dependent terms.

Finally, the relevant quantity needed for the extraction
of the transverse nucleon structure function from $g_T^D$
is plotted in Fig.\ref{F_ET} (solid curves).
The ratio $g_T^D/g_T^N$ comes out to be very similar to the
one obtained from the full convolution model (dashed curves).
In Fig.\ref{F_ET} curves (i) have $\overline{g}_2^N = 0$,
while curves (ii) include the twist-3 component.
As a reference point, the result with a constant depolarization
factor, $(1 - 3/2\,P_D)$, is also shown (dotted line).
These results indicate that the expansion formula
(\ref{expT}) is quite a good approximation to the
convolution model (\ref{conT}) over nearly the entire
$x$-domain of current experiments.

\section{Conclusion}
\label{finita}

We have presented a formulation of spin-dependent deep-inelastic
scattering from spin 1/2 and 1 nuclear targets.
Starting from a covariant framework we have derived
non-relativistic convolution formulae for the nuclear
$g_1^A$ and $g_T^A (\equiv g_1^A + g_2^A)$ structure functions
in terms of polarized nucleon distribution functions and
off-mass-shell extrapolations of the nucleon structure functions
$g_{1,T}^N$.
It is known that relativistically the factorization of
nuclear and nucleon parts of the total structure function,
which is necessary for convolution, does not hold.
Our results, however, are self-consistent to order ${\bf p}^2/M^2$
in the nucleon momentum, and represent the first systematic
derivation of convolution formulae for polarized
structure functions of weakly bound nuclei.
To this order, while $g_1^A$ can be expressed in terms of
$g_1^N$, we find that $g_2^A$ receives contributions from
both the $g_2^N$ {\em and} $g_1^N$ structure functions
of the nucleon.

We have further utilized the fact that the nucleon momentum
distributions in non-relativistic nuclei are strongly peaked
around the light-cone momentum fraction $y \sim 1$ and the
on-mass-shell point $p^2 \sim M^2$.
Expanding the virtual nucleon structure functions
about these points we obtained simple expansion formulae
for $g_{1,T}^A$ valid for $x \alt  0.7$.

The performance of the convolution and expansion approaches
was examined for the case of the deuteron, which has direct
practical implications for the extraction of the free neutron
structure function from deuterium data.
For the $g_1^D$ structure function good agreement was found
between the expansion approximation and the non-relativistic
convolution for all $x$ below $\sim 0.6$.
Differences between the non-relativistic convolution and
previous relativistic calculations become noticeable only
for $x \agt 0.7$.
At smaller $x$ the source of the largest uncertainty is
the non-relativistic deuteron $D$-state probability.

We have also investigated for the first time the nuclear effects
relevant for the extraction of the neutron structure function
$g_T^n$ (or $g_2^n$) from measurements of the transverse structure
function of the deuteron $g_T^D$.
Qualitatively these were found to be similar to the nuclear
effects for $g_1^D$, and for $x \alt 0.5$ largely independent
of the details of the twist-3 component of $g_2^N$.

We can conclude, therefore, that for both $g_1^D$ and $g_T^D$
the non-relativistic expansion formula provides a simple and,
within current experimental accuracy, reliable means of analyzing
nuclear effects in the deuteron.
In future high-precision experiments \cite{SLACE15X,HERMES}
the role of relativistic corrections as well as corrections
to the impulse approximation itself may be more significant.
These experiments should provide us with valuable guidance
as to the relevance of relativistic effects in light nuclei,
and where the impulse approximation may break down.

\acknowledgements

We would like to thank A.W.Thomas for a careful reading
of the manuscript.
S.K., W.M. and G.P. would like to thank the ECT$^*$, Trento,
for its hospitality and support during recent visits,
where some of this work was performed.
S.K. thanks the Alexander von Humboldt Foundation for support.

\appendix

\section{Non-relativistic reduction of the nucleon field operator}
\label{nr-reduction}

To derive the relation in Eq.(\ref{N-psi}) between the relativistic
and non-relativistic field operators, we first write the
relativistic nucleon field operator $N$ in terms of upper
and lower components,
$\varphi$ and $\chi$, respectively:
\begin{eqnarray}
\label{N4}
N({\bf r},t)
&=& e^{-iMt}\left( \begin{array}{c}
                   \varphi({\bf r},t)\\
                   \chi({\bf r},t)
                   \end{array}
            \right).
\end{eqnarray}
In the extreme non-relativistic limit ${\bf p}\to 0,\ \chi\to 0$,
and at zeroth order in $|{\bf p}|/M$, the non-relativistic nucleon
field is simply given by the ``large'' upper component $\varphi$.
We require, however, a relation which is valid to order
${\bf p}^2/M^2$.

The equation of motion for a nucleon field in the presence
of a ``potential'' $\widehat{\cal V}$ can be written:
\begin{eqnarray}
\label{DirE}
\left( i\overlay{\slash}\partial - M \right) N
&=& \widehat{\cal V},
\end{eqnarray}
where
\begin{eqnarray}
\label{V}
\widehat{\cal V}
&=& S + \overlay{\slash}V
    + i\gamma_5 P + \gamma_5\overlay{\slash}A
\end{eqnarray}
describes the interaction of a single nucleon with
mesonic fields produced by the surrounding nucleons.
The first two terms in Eq.(\ref{V}) describe the
coupling of a nucleon to scalar and vector meson fields,
$S=g_s\sigma$ and $V_\mu=g_\omega\omega_\mu$, respectively,
while the last two correspond to pseudoscalar (PS)
and pseudovector (PV) $\pi N$-couplings, $P=g_\pi\pi$
and $A_\mu=(g'_\pi/2M)\ \partial_\mu\pi$
(for simplicity we ignore isospin).

Written for the upper and lower components, the Dirac
equation (\ref{DirE}) reads:
\begin{mathletters}
\begin{eqnarray}
\label{phi}
\left( i \partial_0 - S - V_0 - {\mbox{\boldmath$\sigma\cdot A$}}
\right) \varphi\
-\ \left( {\mbox{\boldmath$\Pi\cdot\sigma$}} + iP - A_0
   \right) \chi
&=& 0, \\
\label{chi}
\left( {\mbox{\boldmath$\Pi\cdot\sigma$}} - iP - A_0
\right) \varphi\
-\ \left( i\partial_0 + 2M - S + V_0
        + {\mbox{\boldmath$\sigma\cdot A$}}
   \right) \chi
&=& 0,
\end{eqnarray}
\end{mathletters}%
where {\boldmath$\Pi$}$\equiv {\bf p}-${\boldmath$V$},
and ${\bf p}= -i${\boldmath$\nabla$} is the momentum operator.
In the non-relativistic limit the driving term in
Eq.(\ref{chi}) is $2M$.
Expanding $\chi$ as a series in inverse powers of $M$ gives:
\begin{eqnarray}
\label{chi2}
\chi
&=&
\left( 1 - \frac{i\partial_0 + S - V_0
         - {\mbox{\boldmath$\sigma\cdot A$}}}{2M} + \ldots
\right)
{ \left( {\mbox{\boldmath$\Pi\cdot\sigma$}} - iP - A_0
  \right)\varphi\
\over 2 M },
\end{eqnarray}
where the leading term is of order $|{\bf p}|/M$,
while the next term is of order $|{\bf p}|^3/M^3$.
For weakly bound nuclei, such as ${}^2$H or ${}^3$He,
the interaction $\widehat{\cal V}$ is of the same order
as the kinetic energy, ${\bf p}^2/2M$.
Furthermore, the kinetic and potential energies are
almost equal in magnitude while opposite in sign
\footnote{Note however that in relativistic models
of nuclear matter, such as the Walecka model,
some parts of the interaction $\widehat{\cal V}$
can be large but of opposite sign (e.g. $S$ and $V_0$),
leading to a small overall $\widehat{{\cal V}}$.},
so that one can treat the different parts of the
interaction ($S,\ V_0$ and {\mbox{\boldmath$\sigma\cdot A$}})
as all being of the order ${\bf p}^2/M^2$.
To this order, it is sufficient therefore
to keep only the first term in the first parentheses
in Eq.(\ref{chi2}).

{}From the equations of motion for the mesonic field we observe
that the ratio of the spatial ({\boldmath$V$}) to time ($V_0$)
components of the vector current is
$\left. \left| \bar N{\mbox{\boldmath$\gamma$}}N \right| \right/
               \bar N\gamma_0 N
\sim \left| {\bf p} \right| \left.\right/M$,
so that $|{\mbox{\boldmath$V$}}/V_0|\sim |{\bf p}|/M$.
Furthermore, since the time component of the axial vector
interaction is given by the time derivative of the pion field,
$A_0=(g'_\pi/2M)\ \partial_0 \pi$, it is reasonable to assume that
$A_0 \sim ({\bf p}^2/M^2)P$.
Therefore to order ${\bf p}^2/M^2$ one can neglect
{\boldmath$V$} and $A_0$ in Eq.(\ref{chi2}),
so that the lower component $\chi$ can be written:
\begin{eqnarray}
\label{chi3}
\chi
&=& \frac1{2M}\left( {\mbox{\boldmath$\sigma\cdot$}}{\bf p} - iP
              \right) \varphi.
\end{eqnarray}
Substituting Eq.(\ref{chi3}) into Eq.(\ref{phi}) leads then
to the Pauli--Schr\"odinger equation for the two-component
nucleon spinor:
\begin{eqnarray}
i\partial_0 \varphi
&=& \left( \frac{{\bf p}^2}{2M} + V_{\text{NR}}
    \right) \varphi,                            \label{SchE} \\
V_{\text{NR}}
&=&
S\ +\ V_0\ +\ {\mbox{\boldmath$\sigma\cdot A$}}\
   +\ \frac{1}{2M}
      \mbox{\boldmath$\sigma\cdot$}(\mbox{\boldmath$\nabla$}P)\
   +\ \frac{P^2}{2M}\ ,                         \label{VNR}
\end{eqnarray}
where $V_{\text{NR}}$ is the non-relativistic analog
of $\widehat{\cal V}$.
Recalling that we assume all parts of the interaction being of the
same order as the kinetic energy, ${\bf p}^2/2M$,
we observe from Eq.(\ref{VNR}) that the PS term should
be of order $P \sim {\mbox{\boldmath$\sigma\cdot$}}{\bf p}$,
and so must be kept in Eq.(\ref{chi3}).

Note also that to order ${\bf p}^2/M^2$ the nucleon density
$N^\dagger N$ receives contributions from the lower component,
$\chi$, in which case $\varphi$ cannot be identified with
the properly normalized non-relativistic nucleon field $\psi$.
Following Ref.\cite{LL}, we introduce a renormalization constant
$Z$ such that $\varphi=Z\psi$, with $Z$ determined by the
particle number (charge) conservation condition:
\begin{eqnarray}
\label{renorm}
\int d^3{\bf r}\, \psi^\dagger\psi
&=&
\int d^3{\bf r}\, N^\dagger N\ .
\end{eqnarray}
Inserting $\chi$ in Eq.(\ref{chi3}) into (\ref{renorm})
we find, to order ${\bf p}^2/M^2$:
\begin{eqnarray}
\label{Z}
Z
&=& 1 -
\frac{1}{8M^2}
\left( {\bf p}^2\
    +\ {\mbox{\boldmath$\sigma\cdot$}}({\mbox{\boldmath$\nabla$}}P)\
    +\ P^2
\right) , \\
N({\bf r},t)
&=& e^{-iMt}
\left(
\begin{array}{r}
Z\, \psi({\bf r},t) \\
{ \left( {\mbox{\boldmath$\sigma\cdot$}\displaystyle{\bf p} - iP}
  \right) \over \displaystyle 2M}\,
\psi({\bf r},t)
\end{array}
\right).
\end{eqnarray}
Therefore the renormalization constant depends explicitly
on the PS pion--nucleon interaction.
In the non-relativistic limit the PV and PS couplings
result in identical $P$-wave pion-nucleon interactions,
as seen from Eq.(\ref{VNR}).
However, the PS coupling also generates a strong $S-$wave
$\pi N$ interaction (the term $P^2/2M$ in Eq.(\ref{VNR})).
Considered alone, the PS term leads to incorrect $\pi N$
scattering lengths, and its contribution is only cancelled
by the introduction of a non-linear $\sigma\pi\pi$ coupling.
In our model therefore we consider only the PV coupling,
which does not lead to the spurious $S-$wave interaction.
In this case the non-relativistic expression for the
four-component nucleon spinor is given by Eq.(\ref{N-psi}),
and the renormalization constant is interaction-independent.

\section{Deuteron Identities}
\label{DI}

For completeness we present here some definitions and
useful identities for the deuteron which are used in
Section~\ref{sf-input}.

The deuteron wavefunction in momentum space is defined as:
\begin{eqnarray}
\label{PsiD}
\Psi_{m}({\bf p})
= \frac{\sqrt{2\pi^2}}p
  \left( u(p)
       - w(p)\, \frac{S_{12}({\bf\widehat p})}{\sqrt8}
  \right)\,\chi_{1,m} ,
\end{eqnarray}
where $p=|{\bf p}|$ and $m$ is the projection of the deuteron
spin on the axis of quantization,
$\chi_{1,m}$ is the spin 1 wavefunction of the two-nucleon system,
and $S_{12}({\bf\widehat p})$ is the tensor operator with
${\bf\widehat p}={\bf p}/|{\bf p}|$.
We define the $S$- and $D$-state wavefunctions in momentum space as:
\begin{mathletters}
\begin{eqnarray}
\label{uw}
\sqrt{2\pi^2}\,u(p)
&=& \int_0^{\infty}dr\,pr\,j_0(pr)\,u(r) , \\
\sqrt{2\pi^2}\,w(p)
&=& \int_0^{\infty}dr\,pr\,j_2(pr)\,w(r) ,
\end{eqnarray}
\end{mathletters}%
where $u(r)$ and $w(r)$ are the standard wavefunctions
in configuration space,
and normalized such that:
\begin{equation}
\int_0^\infty dp\,\left( u^2(p)+w^2(p)\right) = 1.
\end{equation}

If ${\bf S}$ is a unit vector along the direction of the
spin quantization, then
\begin{eqnarray}
{1\over 2}
\chi^{\dagger}_{1,m}
\left( {\mbox{\boldmath$\sigma$}}^{(p)}
     + {\mbox{\boldmath$\sigma$}}^{(n)}
\right)
\chi_{1,m}
&=& m {\bf S},
\end{eqnarray}
where ${\mbox{\boldmath$\sigma$}}^{(p),(n)}$ are SU(2) Pauli
spin matrices acting on the proton and neutron wave function
respectively.
Using some simple relations from the SU(2) algebra
we find:
\begin{eqnarray}
\Psi^{\dagger}_m({\bf p})
\left( {\mbox{\boldmath$\sigma$}}^{(p)}
     + {\mbox{\boldmath$\sigma$}}^{(n)}
\right)
\Psi_{m}({\bf p})
&=&
m\, \frac{4\pi^2}{{\bf p}^2}
\left[ {\bf S}
       \left( u^2 + \frac{uw}{\sqrt2} - w^2 \right)
+  {3\over 2} \widehat{{\bf p}}
    \left( \widehat{{\bf p}} \cdot {\bf S} \right)
    w \left( w - \sqrt{2} u \right)
\right].
\label{genME}
\end{eqnarray}


\newpage
\begin{figure}
\caption{Deep inelastic scattering from a polarized nucleus
        in the impulse approximation.
        The momenta of the target nucleus ($P$),
        virtual nucleon ($p$)
        and photon ($q$) are marked, and $S$ denotes the nuclear
        spin vector.}
\label{F_DIS}   
\end{figure}

\begin{figure}  
\caption{Nucleon structure function input used in the calculation
        of nuclear structure functions, for $Q^2 = 4$ GeV$^2$:
    (a) parametrization \protect\cite{GS} of the proton
        \protect\cite{EMC,SMCP,SLAC}
        and neutron \protect\cite{E143P} $xg_1$ data;
    (b) isoscalar nucleon structure functions
        $xg_1^N \left(= (xg_1^p + xg_1^n)/2\right)$ (solid),
        $xg_2^N$ (dashed),
        and $xg_T^N \left(= xg_1^N + xg_2^N\right)$ (dotted).
        In (b) curves (1) contain the twist-2 component of $g_2^N$
        only, while in curves (2) $g_2^N$ has in addition a twist-3
        contribution based on the bag model calculation of
        Ref.\protect\cite{STRAT}.}
\label{F_NSF}   
\end{figure}

\begin{figure}  
\caption{Ratio of the deuteron to nucleon $g_1$ structure functions
        calculated via the convolution formula in
        Eq.(\protect\ref{con1}), with the Paris \protect\cite{PARIS}
        (solid), Bonn \protect\cite{BONN} (dotted) and Gross
        \protect\cite{BG} (dashed) deuteron wavefunctions.
        The latter, which also include small $P$-state components,
        are renormalized so that the $S$- and $D$-state
        wavefunctions alone are normalized to unity.}
\label{F_C1}    
\end{figure}

\begin{figure}  
\caption{Ratio of the transverse $g_T$ deuteron to nucleon structure
        functions within the convolution model,
        Eq.(\protect\ref{conT}),
        for the Paris wavefunction \protect\cite{PARIS},
        with (solid) and without (dotted)
        the twist-3 component of $g_2^N$.
        For comparison the $g_1^D/g_1^N$ ratio (dashed) from
        Fig.\protect\ref{F_C1} is also shown.}
\label{F_CT}    
\end{figure}

\begin{figure}  
\caption{Contributions to the deuteron $xg_1^D$ structure function
        from various terms in the expansion formula,
        Eq.(\protect\ref{exp1}):
        zeroth order term (solid), first derivative (dashed),
        second derivative (dot-dashed),
        nucleon off-shell contribution (dotted).
        The latter three higher order terms are scaled
        by a factor 100.}
\label{F_TERMS1} 
\end{figure}

\begin{figure}  
\caption{Deuteron to nucleon structure function ratio,
        $g_1^D/g_1^N$, using the expansion formula (solid),
        compared with the convolution result (dashed) from
        Fig.\protect\ref{F_C1}, and with a constant depolarization
        factor $(1-3/2\,P_D)$ (dotted), with $P_D \simeq 5.8\%$ from
        the Paris wavefunction \protect\cite{PARIS}.}
\label{F_E1}    
\end{figure}

\begin{figure}  
\caption{Deuteron to nucleon $g_1$ structure function ratio from
        the non-relativistic expansion formula with (solid)
        and without (dashed) the off-shell component in
        Eq.(\protect\ref{exp1}),
        for the deuteron wavefunction of Ref.\protect\cite{BG}.
        Shown also is the result of the relativistic
        calculation of Ref.\protect\cite{MPT} (dotted).}
\label{F_E1OFF}    
\end{figure}

\begin{figure}  
\caption{Contributions to the $xg_T^D$ structure function
        in the expansion formula, Eq.(\protect\ref{expT}):
    (a) $xg_T^N$ components --- zeroth order term (solid),
        first (dashed)
        and second (dot-dashed) derivatives;
    (b) $xg_2^N$ (triple-dot--dashed)
        and $(xg_2^N)'$ (dashed) components,
        on the background of the zeroth order term (solid)
        in (a).}
\label{F_TERMST} 
\end{figure}

\begin{figure}  
\caption{Transverse deuteron to nucleon structure function ratio
        using the expansion formula (solid),
        Eq.(\protect\ref{expT}),
        compared with the convolution model result (dashed).
        Curves (i) have $\overline{g}_2^N = 0$, while curves (ii)
        include the twist-3 component.
        The dotted curve indicates the constant depolarization
        ratio, $(1 - 3/2\,P_D)$, with $P_D \simeq 5.8\%$ from
        the Paris potential \protect\cite{PARIS}.}
\label{F_ET}    
\end{figure}

\end{document}